# Direct growth of crystalline spinel γ-GaON films: A wide-bandgap mixed-anion semiconductor

Baikui Li[1*], Zecheng Yang[1], Wenhao Yin[1], Zijing Wu[1], Haitao Zhao[1], Yunwei Sheng[2], Honglei Wu[1], and Jiannong Wang[2]

[1]*State Key Laboratory of Radio Frequency Heterogeneous Integration, College of Physics and Optoelectronic Engineering, Shenzhen University, 3688 Nanhai Avenue, Shenzhen, China*

[2]*Physics Department, The Hong Kong University of Science and Technology, Clear Water Bay, Kowloon, Hong Kong, China*

**Abstract**

Mixed-anion semiconductors offer opportunities for bandgap engineering beyond conventional multication alloys, but competing phases and disorder complicate their synthesis and the identification of their electronic properties. Gallium oxynitride (GaON) could connect the technologically important GaN and $Ga_2O_3$ systems, yet establishing a clear relation between its crystal structure, composition and bandgap remains challenging. Here we report the direct growth of crystalline spinel γ-GaON films by nitrogen-plasma-assisted pulsed-laser deposition. Varying the nitrogen/oxygen supply yields monoclinic β-$Ga_2O_3$, predominantly disordered films and crystalline spinel GaON. Complementary diffraction, atomic-resolution microscopy and elemental analysis establish the spinel structure and nitrogen incorporation. Temperature-dependent photocurrent spectroscopy reveals a clear redshift of the crystalline γ-GaON response, supporting assignment of a band-edge transition near 4.3 eV at room temperature. These results open opportunities to investigate mixed-anion wide-bandgap semiconductor physics and develop heterostructures.

## Introduction

Composition engineering underpins semiconductor technologies ranging from light-emitting devices [1–8] to high-electron-mobility transistors [9–12]. Much of this progress has relied on mixing cations to tailor band structures and functional properties. Mixed-anion chemistry offers a complementary, less explored approach to semiconductor design by incorporating different anions into a common crystalline framework. However, realizing crystalline mixed-anion compounds remains challenging because anion incorporation can be accompanied by competing phases and structural disorder. Establishing how anion composition modifies semiconductor properties therefore requires controlled synthesis together with reliable structural, chemical and optical characterization.

Gallium oxynitride (GaON) is a candidate for such mixed-anion semiconductor research. Combining oxygen and nitrogen within a gallium-based compound could connect the technologically important GaN and $Ga_2O_3$ material systems and provide additional possibilities for engineering semiconductor properties. A central experimental challenge is to obtain crystalline GaON in a form that permits its structure, composition and electronic response to be examined together and supports fabrication of devices in which GaON serves as the active semiconductor.

Extensive efforts have been devoted to understanding and synthesizing GaON over the past several decades [17–44]. Early studies suggested the formation of gallium oxynitride through reactions of ammonia with gallium oxides and through vapour-phase synthesis [17,18], and investigated GaON films for GaAs surface passivation [19]. Subsequent surface-sensitive studies examined Ga–O–N formation during oxidation of GaN [20–22]. A metastable nanocrystalline cubic GaON phase was later observed at an intermediate stage of precursor-derived GaN formation [23], followed by computational investigations of the structure and electronic properties of spinel gallium oxynitrides [24–26]. Micrometre-scale crystalline GaON with spinel-type structures was subsequently synthesized under high-pressure, high-temperature conditions [27–30]. Two contemporaneous reports described spinel microcrystals obtained at 7 GPa and 1100 °C [27] and at 5 GPa and 1700 °C [28], establishing a foundation for investigating γ-GaON as a distinct crystalline compound.

Later efforts extended GaON synthesis to nanostructures [31–35] and ultrathin crystalline layers on GaN [36–38] and ZnO surfaces [39]. GaN-derived surface layers were investigated as channel or interfacial components in GaN-based devices [36–38]. Meanwhile, sputtering [40–42] and atomic layer deposition (ALD) [43,44] enabled preparation of Ga–O–N films with varied compositions and optical responses, including amorphous GaON. Annealing-induced spinel formation was also reported in sputtered films at 800 °C [41], although the γ-GaON assignment relied on diffraction and optically inferred composition without independent elemental verification. Together, these studies demonstrated the chemical versatility and technological potential of the Ga–O–N system, with applications ranging from photocatalysis

[31,34,35,39,45] to photodetection [40,42,44,46,47] and GaN-based power electronic devices [36,38,48].

Despite these advances, the available material forms have constrained investigation of GaON as a semiconductor in its own right. High-pressure, high-temperature synthesis has yielded micrometre-scale crystals, limiting the dimensions available for device fabrication and systematic property measurements [27–30]. Amorphous films and nanostructured materials introduce disorder, defects and surface contributions that complicate the extraction of intrinsic properties [31–35,40–44]. Ultrathin crystalline layers have demonstrated benefits in existing semiconductor structures [36–39], but leave the behaviour of GaON as an independently characterized active semiconductor less fully explored. Reported structures include monoclinic, wurtzite, spinel and amorphous forms [21,24–44,49], while reported optical gaps span approximately 2.2–4.9 eV [21,24,28,31–37,41–44,49]. This variation reflects differences in composition, phase and disorder, as well as optical assignment methods. Access to crystalline GaON with dimensions suitable for device fabrication, together with reliable identification of its band-edge response, is therefore essential for exploring its electronic and optoelectronic potential.

Here we report direct growth of crystalline spinel γ-GaON films by nitrogen-plasma-assisted pulsed-laser deposition (NPA-PLD). The representative film has a total thickness of approximately 234 nm, with γ-GaON overlying a thin β-phase interfacial region. Varying the nitrogen/oxygen supply reveals distinct structural regimes and a non-monotonic evolution of crystalline order. Diffraction, atomic-resolution microscopy and nitrogen-sensitive spectroscopy jointly substantiate the γ-GaON identification. Temperature-dependent photocurrent spectroscopy, interpreted alongside the β-$Ga_2O_3$ reference and the contrasting response of a predominantly disordered film, supports assignment of a transition near 4.3 eV to the γ-GaON band edge. These results establish a direct growth route to crystalline GaON films and connect their structural and chemical identification with a band-edge-related optical response, providing a basis for investigating GaON semiconductor properties and devices.

## Results

### Growth-dependent structural evolution of Ga–O–N films

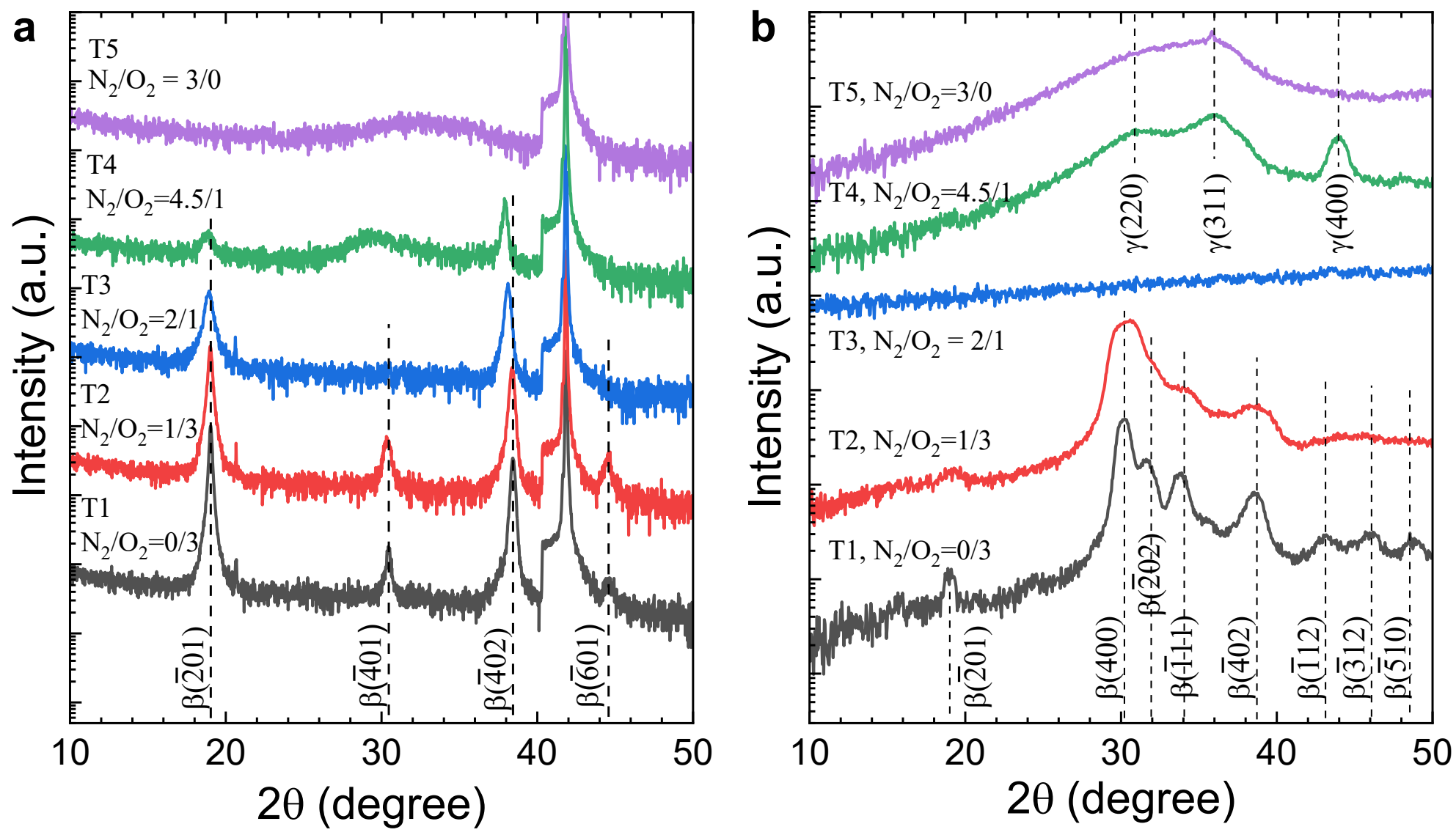

**Fig. 1.** Growth-dependent structural evolution of Ga–O–N films. **a**, X-ray diffraction (XRD) patterns and **b**, grazing-incidence wide-angle X-ray scattering (GIWAXS) patterns of films deposited at 550 °C and 1 Pa. T1–T5 correspond to $N_2/O_2$ flow settings of 0/3, 1/3, 2/1, 4.5/1 and 3/0 sccm, respectively. Across these conditions, the dominant diffraction signatures change from β-$Ga_2O_3$ to a predominantly disordered film and then to a spinel phase. The spinel reflections broaden under $N_2$-only growth. These are comparisons between separately grown films.

Ga–O–N films were deposited from a β-$Ga_2O_3$ target using NPA-PLD with independently supplied $N_2$ and $O_2$ (Supplementary Fig. S1). Five films, T1–T5, were grown using $N_2/O_2$ flow settings of 0/3, 1/3, 2/1, 4.5/1 and 3/0 sccm, respectively. The corresponding $N_2/(N_2+O_2)$ flow fractions are 0, 0.25, approximately 0.67, approximately 0.82 and 1. This fraction specifies the supplied gases; it is not a measurement of nitrogen chemical potential or film composition. The total gas flow also differs between conditions. Film thickness and surface morphology are summarized in Supplementary Fig. S2, and the diffraction patterns are shown in Fig. 1.

For T1, grown without nitrogen plasma, the film reflections are consistent with monoclinic β-$Ga_2O_3$ with a preferred (400) orientation (Fig. 1a,b). The β-phase reflections remain in T2 but broaden after nitrogen is introduced into the growth atmosphere. This change indicates reduced crystalline coherence and/or increased structural disorder under the modified deposition conditions; diffraction alone does not quantify nitrogen incorporation.

At an $N_2/O_2$ setting of 2/1 (T3), the β-$Ga_2O_3$ reflections are strongly suppressed and the GIWAXS pattern is nearly featureless. Weak β-$Ga_2O_3$ peaks remain detectable by XRD. We therefore describe T3 as predominantly disordered, with residual crystalline β-$Ga_2O_3$, rather than as a completely amorphous single-phase material.

At 4.5/1 (T4), reflections at 30.96°, 35.94° and 43.98° are consistent with the (220), (311) and (400) planes of a spinel lattice [27–30,41,50]. These reflections dominate the GIWAXS pattern, although weak β-$Ga_2O_3$ peaks remain in XRD. The data establish the formation of a spinel-containing film with residual β phase; they do not establish phase purity. Microscopy and elemental analyses below address the chemical identity of the spinel phase.

Under $N_2$-only growth (T5), the spinel reflections persist but broaden, and higher-order reflections are no longer resolved. The crystalline order therefore does not improve monotonically with the $N_2$ flow fraction. Among the five conditions examined, T4 gives the clearest spinel signature. This series identifies conditions favorable to spinel formation, without defining precise growth-window boundaries. Because T1–T5 were grown separately, their structural sequence does not establish a temporal amorphous-to-spinel transformation within an individual film.

### Identification of crystalline γ-GaON

To identify the atomic structure and chemical nature of the spinel phase observed in Fig.1, cross-sectional scanning transmission electron microscopy (STEM), electron energy-loss spectroscopy (EELS) and time-of-flight secondary ion mass spectrometry (ToF-SIMS) measurements were performed on films T1 and T4, which exhibit representative β-$Ga_2O_3$ and spinel diffraction characteristics, respectively.

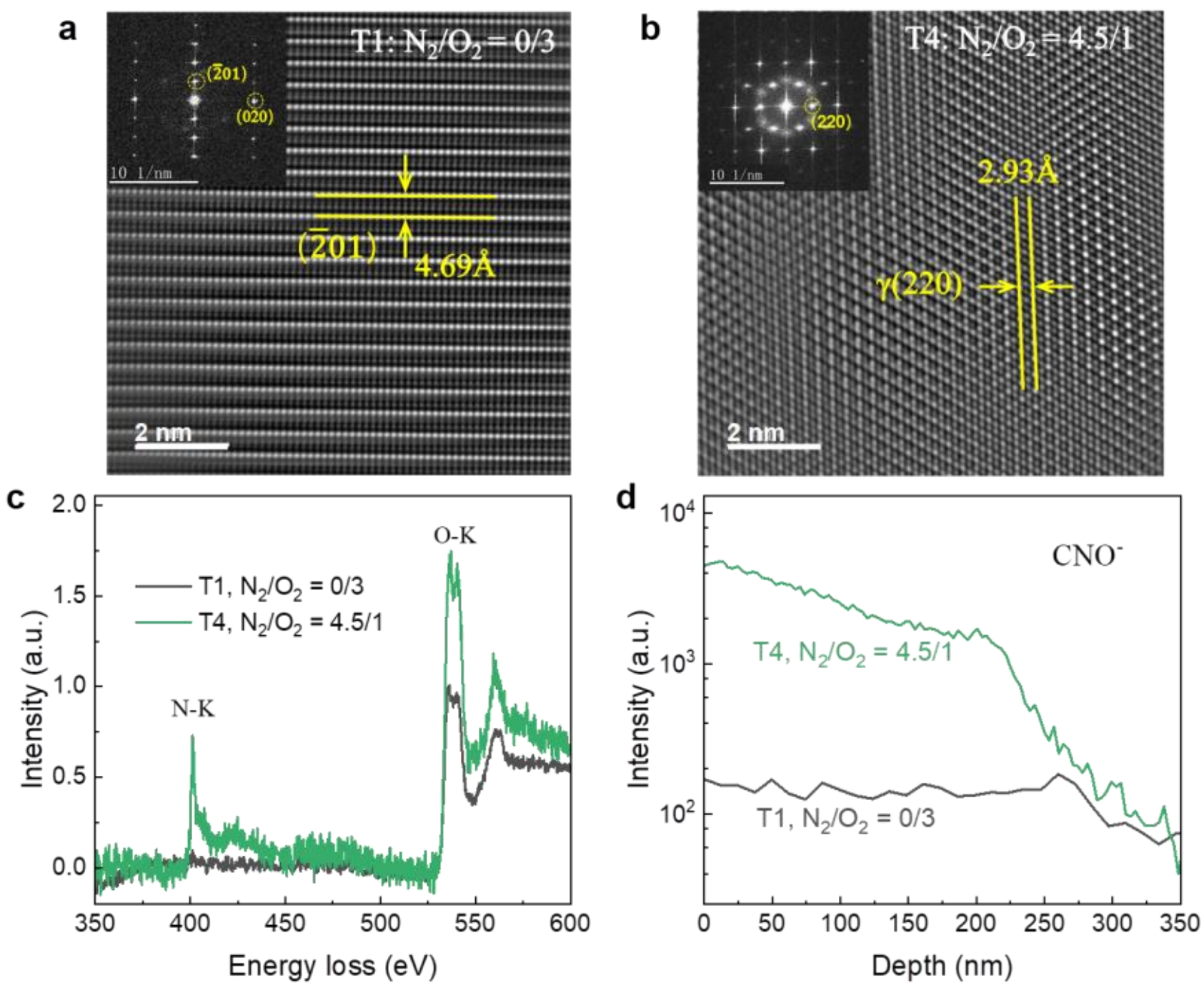

**Fig. 2.** Structural and compositional identification of γ-GaON. **a, b,** Cross-sectional HAADF-STEM images of T1 and T4 showing spacings of 4.69 Å and 2.93 Å, assigned to β-$Ga_2O_3$ (−201) and spinel (220), respectively. Insets show the corresponding FFT patterns. **c,** EELS spectra after background subtraction: T1 shows the O-K edge, whereas T4 shows both O-K and N-K edges. **d,** ToF-SIMS $CNO^-$ depth profiles, showing a stronger nitrogen-related signal in T4. The combined measurements support identification of the nitrogen-containing spinel phase in T4 as γ-GaON.

The T1 HAADF-STEM image contains lattice fringes with a spacing of 4.69 Å, consistent with the (−201) plane of $\beta$-$Ga_2O_3$ (Fig. 2a). T4 contains a distinct lattice arrangement with a spacing of 2.93 Å, consistent with a spinel (220) plane (Fig. 2b). This spacing and the FFT pattern support the structural assignment from diffraction. Because an oxide spinel can exhibit a similar spacing, elemental measurements are needed to identify the oxynitride.

EELS acquired during STEM characterization detects both oxygen and nitrogen in T4. Following power-law background subtraction, T1 shows the O-K edge near 532 eV, whereas T4 additionally shows the N-K edge near 401 eV (Fig. 2c). These measurements, together with the spinel lattice observations, support assignment to crystalline γ-GaON. Additional HRTEM images show a β-phase interfacial region approximately 15 nm thick beneath the predominantly γ-GaON region (Supplementary Fig. S3). Spatially resolved EELS further shows a weaker N-K signal near the substrate (Supplementary Fig. S4). The 234-nm total thickness of T4 therefore does not represent a phase-pure γ-GaON layer.

ToF-SIMS provides complementary evidence for nitrogen incorporation (Fig. 2d). The $CNO^-$ signal is substantially stronger in T4 than in the nitrogen-free-growth reference T1 and decreases towards the substrate. This trend is qualitatively consistent with the EELS depth dependence. The $CNO^-$ intensity is a secondary-ion signal rather than a calibrated nitrogen concentration, and is used here as supporting evidence for the nitrogen distribution.

An ideal $Ga_3O_3N$ composition has an N/O ratio of 1/3. Integration of the N-K and O-K edges over identical 50-eV windows, followed by correction for the relative ionization cross sections, gives an estimated N/O ratio of approximately 0.23 for the analyzed T4 region (Supplementary Fig. S4). With oxygen normalized to three and Ga assumed to be trivalent, charge balance gives the nominal formula $Ga_{2.69}O_3N_{0.69}$. This is a semiquantitative compositional representation, not a refinement of site occupancies or a measurement of uniform stoichiometry throughout the film. We use γ-GaON to denote the crystalline nitrogen-containing spinel phase without implying the ideal $Ga_3O_3N$ composition.

Taken together, the diffraction patterns, lattice images and nitrogen-sensitive measurements identify γ-GaON in the directly deposited T4 film. The material combines crystalline spinel regions with a residual β-phase interface and a depth-dependent nitrogen signal. These observations establish a film platform for examining GaON properties while retaining the structural and compositional qualifications needed to interpret its optical response.

**Temperature-dependent optical response of crystalline γ-GaON**

Determining an optical gap in GaON requires separating phase-dependent electronic transitions from responses associated with disorder and defects. The transmittance-derived Tauc plots for T1–T5 show curved regions and fitting-interval-dependent intercepts (Supplementary Fig. S5), preventing a unique gap assignment from these plots alone. We therefore examine the temperature dependence of the photocurrent together with the structural evidence.

Temperature-dependent photocurrent spectroscopy supplies an additional discriminator between spectral features. Band-edge energies can shift with temperature, while the response of a device also depends on carrier trapping and collection. Consequently, a redshift supports a band-edge interpretation but is not, by itself, a unique test of an intrinsic transition. Here the comparison with β-$Ga_2O_3$ reference T1 and the contrasting structures of T3 and T4 provide the basis for the assignments.

Normalized photocurrent spectra of metal–semiconductor–metal (MSM) devices on T1–T5 were measured between 25 and 250 °C (Fig. 3a). The dominant peaks in T1, T2 and T4 redshift appreciably, whereas those in T3 and T5 are less temperature dependent. Normalization enables comparison of spectral positions and shapes; it does not establish differences in absolute responsivity or carrier collection efficiency.

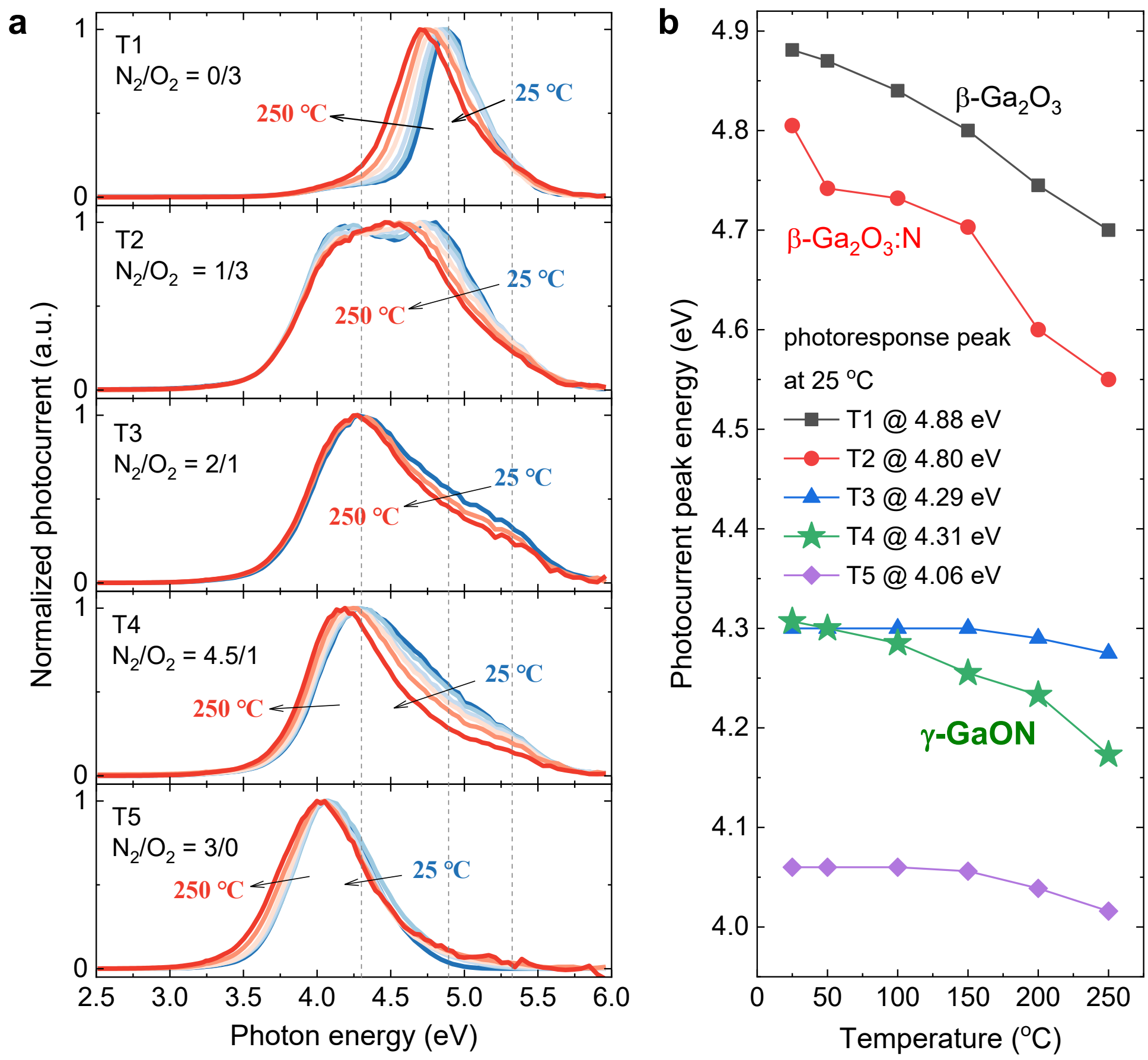

**Fig. 3.** Temperature-dependent photocurrent spectra and peak energies. **a**, Normalized photocurrent spectra of MSM devices on T1–T5, from top to bottom, measured between 25 and 250 °C. **b**, Temperature dependence of the dominant photocurrent peak energies. T1, T2 and T4 show pronounced redshifts. T3 and T4 have similar peak energies near 4.3 eV at room temperature but different temperature dependences. Together with the structural evidence, the T4 response supports assignment to a γ-GaON band-edge transition. The plotted quantities are photocurrent peak energies, not independent measurements of the fundamental gap for every sample.

For $\beta$-$Ga_2O_3$ reference T1, the dominant peak shifts from 4.88 eV at room temperature to approximately 4.70 eV at 250 °C, consistent with a $\beta$-$Ga_2O_3$ band-edge response [50]. T2 has a high-energy peak near 4.80 eV at room temperature that similarly redshifts, suggesting a modified β-phase band-edge response under nitrogen-containing growth conditions. A second response near 4.2 eV is weakly temperature dependent and is consistent with a defect-related contribution. The present measurements do not identify a specific microscopic defect.

T3 exhibits a dominant peak at approximately 4.29 eV with little temperature dependence. Its predominantly disordered structure and this spectral behavior favor an interpretation involving disorder- or defect-related states rather than the same band-edge transition observed in crystalline T4. Weaker high-energy responses near 4.88 and 5.3 eV redshift with temperature. The former is consistent with residual $\beta$-$Ga_2O_3$; the latter is not assigned uniquely, because spectral energy alone is insufficient to identify an additional crystalline phase. A fundamental gap for the disordered component of T3 cannot be extracted reliably from these data.

In T4, the dominant peak shifts from 4.31 eV at 25 °C to approximately 4.17 eV at 250 °C. The pronounced redshift, together with the spinel structure and nitrogen incorporation, supports assignment of this response to a band-edge transition of γ-GaON. We therefore identify an optical transition near 4.3 eV associated with the band edge of the investigated film. Its peak position is not assumed to be an exact, composition-independent fundamental gap: T4 is non-stoichiometric and contains an interfacial β phase, and photocurrent depends on both optical excitation and carrier collection. Higher-energy responses are consistent with the residual $\beta$-$Ga_2O_3$ identified structurally.

T5 shows a dominant peak near 4.06 eV with a comparatively weak temperature dependence. Together with the broadened spinel diffraction peaks, this behavior is consistent with an increased contribution from disorder or defects. The lower peak energy cannot therefore be taken as evidence of continuous bandgap tuning by nitrogen content. The data do not permit an unambiguous γ-GaON bandgap assignment for T5.

The comparison between T3 and T4 is particularly informative: nearly coincident room-temperature peak energies accompany different crystalline structures and different thermal shifts (Fig. 3b). Thus, a response near 4.3 eV alone is insufficient to identify a γ-GaON band edge. The combination of diffraction, microscopy, elemental analysis and temperature-dependent spectroscopy provides the stronger basis for assigning the T4 response.

After annealing T4 in $N_2$ for 1 h, γ-GaON diffraction features remain observable at 700 and 800 °C (Supplementary Fig. S6). They weaken and broaden at 900 °C, and prominent $\beta$-$Ga_2O_3$ reflections emerge at 1000 °C. These results establish retention of the spinel signature under the specified annealing conditions and structural conversion at higher temperatures; they do not establish unchanged nitrogen composition or electronic properties after annealing.

## Discussion

Fig. 4 summarizes an empirical relationship between growth conditions, structure and optical response in Ga–O–N films. At fixed substrate temperature and chamber pressure, varying the supplied gases changes both the dominant phase and the degree of crystalline order. The map brings the structural and spectroscopic observations together without treating every photocurrent peak as a fundamental bandgap.

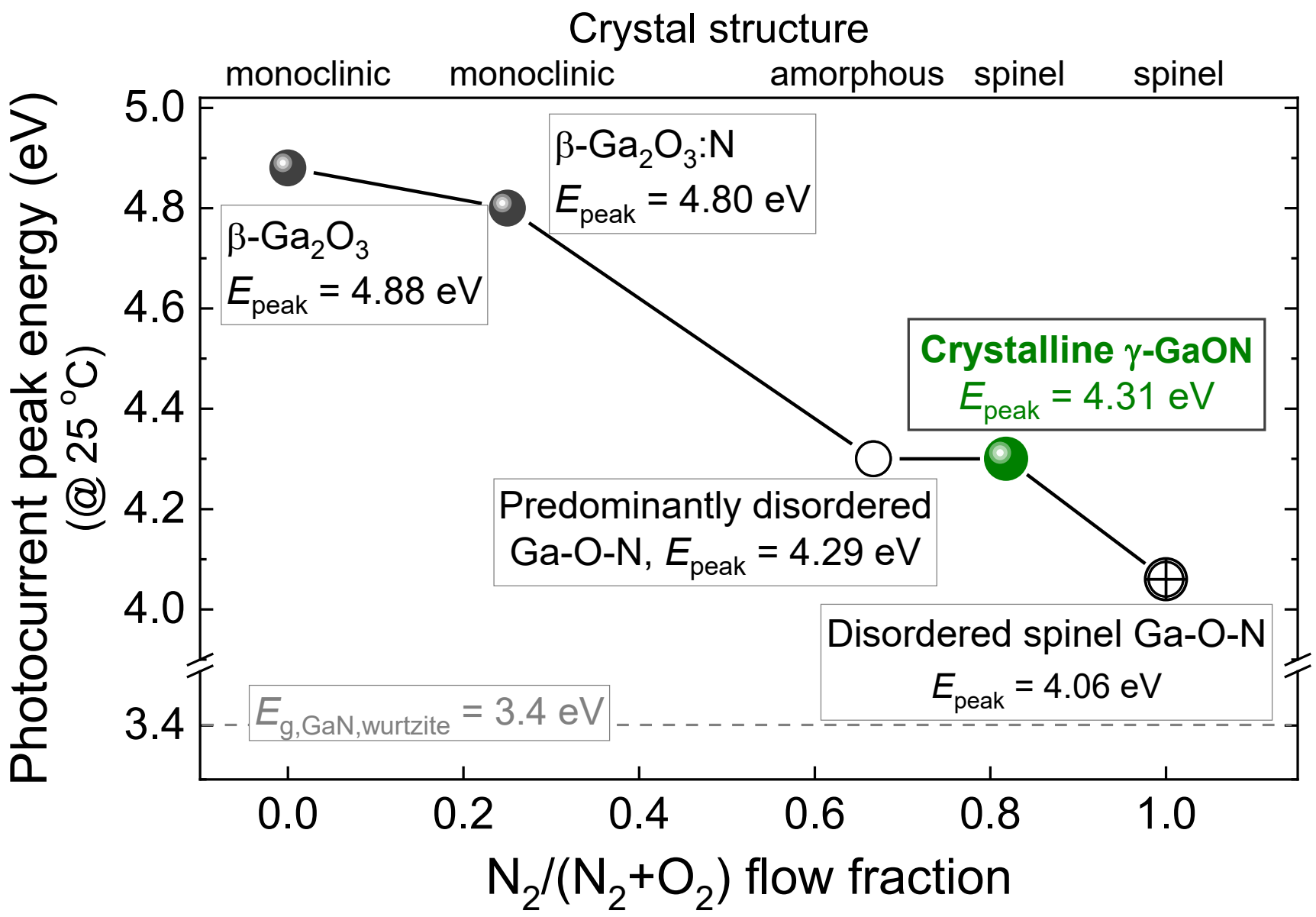

**Fig. 4.** Empirical growth–structure–optical-response map of Ga–O–N films. Dominant structural signatures and room-temperature photocurrent peak energies for T1–T5 are arranged by the supplied $N_2/(N_2+O_2)$ flow fraction. The responses of T1, T2 and T4 are assigned to band-edge-related transitions, whereas those of T3 and T5 are interpreted as predominantly disorder- or defect-related. Connecting lines guide the eye between separate samples; they do not establish a continuous gap-composition relation or phase boundaries. The flow fraction is an experimental control variable, not a calibrated chemical potential or nitrogen atomic fraction.

The synthesis advance is direct growth of crystalline spinel GaON at 550 °C and 1 Pa without post-deposition annealing, supported by complementary structural and nitrogen-sensitive measurements. Representative microcrystal syntheses required 7 GPa at 1100 °C [27] or 5 GPa at 1700 °C [28]. The present route provides crystalline GaON in a deposited film geometry suitable for integrated structural, chemical and optical investigation. The approximately 234-nm total thickness of T4 extends beyond previously reported ultrathin surface and interfacial geometries [36–39], while including a residual β-phase interfacial region. This material form enables examination of GaON semiconductor behaviour and provides a starting point for subsequent control of film dimensions and composition.

A second result is the non-monotonic relationship between the supplied $N_2$ fraction and crystalline order. T4 has a clearer spinel signature than either the predominantly disordered T3 or the $N_2$-only-grown T5. This indicates that increasing the nitrogen fraction in the growth gas does not necessarily improve crystallization. The five conditions establish this trend, but do not isolate the roles of nitrogen incorporation, oxygen supply and total flow, or determine a microscopic growth mechanism. The association between a nitrogen-poor β-phase interface and the overlying γ-GaON region provides additional evidence of spatially varying phase formation within T4.

The optical measurements show why this structural information matters. Similar room-temperature peak energies in T3 and T4 do not imply the same electronic transition. Their contrasting temperature dependences, combined with the structural and compositional evidence, support a γ-GaON band-edge assignment near 4.3 eV in T4. This provides an experimentally grounded optical reference for the investigated film, rather than a universal gap for all GaON compositions. Direct growth and correlated characterization together make crystalline GaON accessible for subsequent studies of composition-dependent electronic properties and semiconductor interfaces.

## Methods

### Nitrogen-plasma-assisted pulsed laser deposition of Ga–O–N films

Ga–O–N thin films were deposited on c-plane sapphire substrates by nitrogen-plasma-assisted pulsed laser deposition (NPA-PLD) using a PR-B111 deposition system (Arrayed Materials) equipped with a radio-frequency nitrogen plasma source, independent $N_2$ and $O_2$ gas supply lines, and a COMPex 102F KrF excimer laser (248 nm, Coherent). The substrates were maintained at 550 °C under a chamber pressure of 1 Pa [16]. A β-$Ga_2O_3$ ceramic target (99.99%, MTI Corp.) was ablated using a laser pulse energy of 300 mJ at a repetition rate of 10 Hz. Five representative films were prepared by varying the N2/O2 flow ratio (0/3, 1/3, 2/1, 4.5/1 and 3/0 sccm), denoted T1–T5 throughout this work. The RF plasma power was maintained at 400 W whenever nitrogen plasma was employed. The supplied $N_2$ and $O_2$ flow rates were varied while substrate temperature, chamber pressure, laser settings and deposition duration were held fixed. The total supplied flow was not constant across the series. The $N_2$ flow fraction is used to describe the gas supply, without conversion to a nitrogen chemical potential. Each deposition lasted 16.6 min.

**Structural characterization**

Crystal structures were characterized by X-ray diffraction (XRD, Bruker D8 Advance, Cu Kα radiation) and grazing-incidence wide-angle X-ray scattering (GIWAXS, Xeuss 3.0 UHR, Cu Kα radiation). Cross-sectional atomic structures were examined using an aberration-corrected FEI Titan Cubed Themis G2 300 scanning transmission electron microscope operating in high-angle annular dark-field (HAADF-STEM) mode.

**Chemical characterization**

Local oxygen and nitrogen signals were measured by electron energy-loss spectroscopy (EELS) during STEM characterization. A power-law background was subtracted before comparing the N-K and O-K edges. For the semiquantitative N/O estimate, edge intensities were integrated over 50-eV windows and corrected using the cross-section ratio specified in Supplementary Fig. S4. ToF-SIMS $CNO^-$ depth profiles were used to compare nitrogen-related signals in T1 and T4.

**Device fabrication and photocurrent spectroscopy**

Metal–semiconductor–metal photodetectors were fabricated using Au electrodes deposited on the Ga–O–N films. Temperature-dependent photocurrent spectra were measured using a CS4 optoelectronic probe station equipped with a monochromated EQ-99 broadband light source. The photocurrent was detected using a lock-in amplifier (SR830) and a source-measure unit (Keysight B2912A).

**Data availability**

The data supporting the findings of this study are available within the Article and its Supplementary Information. Additional data are available from the corresponding author upon reasonable request.

**Acknowledgements**

This work was supported by the National Natural Science Foundation of China (62674152, 62074103 and 62474113), the Research Grants Council of the Hong Kong SAR (N_HKUST611/23 and STG3/E-602/23-N), and the State Key Laboratory of Radio Frequency Heterogeneous Integration (Independent Scientific Research Program No. 2024007 and 2026003). We acknowledge technical support from the Electron Microscope Center of Shenzhen University and the Materials Characterization and Preparation Facility (MCPF) of the Hong Kong University of Science and Technology (Guangzhou).

**Author contributions**

B.L. conceived the experiments. Z.Y., B.L. and W.Y. grew the samples and performed relevant measurements. Z.W., H.Z., Y.S. provided support in the measurements. B.L., J.W. and Z.Y. analyzed the data and wrote the manuscript. All authors discussed the results and commented on the manuscript.

**Competing Interests**

The authors declare no competing interests.

**Additional Information**

**Correspondence and requests for materials** should be addressed to Baikui Li (libk@szu.edu.cn).